\input epsf

\newfam\scrfam
\batchmode\font\tenscr=rsfs10 \errorstopmode
\ifx\tenscr\nullfont
        \message{rsfs script font not available. Replacing with calligraphic.}
        \def\scr{\cal}
\else   
        \font\sevenscr=rsfs7
        \font\fivescr=rsfs5
        \skewchar\tenscr='177 \skewchar\sevenscr='177 \skewchar\fivescr='177
        \textfont\scrfam=\tenscr \scriptfont\scrfam=\sevenscr
        \scriptscriptfont\scrfam=\fivescr
        \def\scr{\fam\scrfam}
        \def\cal{\scr}
\fi
\catcode`\@=11
\newfam\frakfam
\batchmode\font\tenfrak=eufm10 \errorstopmode
\ifx\tenfrak\nullfont
        \message{eufm font not available. Replacing with italic.}
        \def\frak{\it}
\else
	
	\font\sevenfrak=eufm7 \font\fivefrak=eufm5
        \font\eightfrak=eufm8
	\textfont\frakfam=\tenfrak
	\scriptfont\frakfam=\sevenfrak \scriptscriptfont\frakfam=\fivefrak
	\def\frak{\fam\frakfam}
\fi
\catcode`\@=\active
\newfam\msbfam
\batchmode\font\twelvemsb=msbm10 scaled\magstep1 \errorstopmode
\ifx\twelvemsb\nullfont\def\Bbb{\bf}
        
	\font\eightbbb=cmb10 at 8pt
	\message{Blackboard bold not available. Replacing with boldface.}
\else   \catcode`\@=11
        \font\tenmsb=msbm10 \font\sevenmsb=msbm7 \font\fivemsb=msbm5
        \textfont\msbfam=\tenmsb
        \scriptfont\msbfam=\sevenmsb \scriptscriptfont\msbfam=\fivemsb
        \def\Bbb{\relax\expandafter\Bbb@}
        \def\Bbb@#1{{\Bbb@@{#1}}}
        \def\Bbb@@#1{\fam\msbfam\relax#1}
        \catcode`\@=\active
	
	\font\eightbbb=msbm8
\fi
        \font\fivemi=cmmi5
        \font\sixmi=cmmi6
        \font\eightrm=cmr8              \def\xrm{\eightrm}
        \font\eightbf=cmbx8             \def\xbf{\eightbf}
        \font\eightit=cmti10 at 8pt     \def\xit{\eightit}
                
        \font\eighttt=cmtt8             
        \font\eightcp=cmcsc8
        \font\eighti=cmmi8              \def\xold{\eighti}
        \font\eightmi=cmmi8
        \font\eightib=cmmib8             \def\xbold{\eightib}
        \font\teni=cmmi10               \def\old{\teni}
        \font\tencp=cmcsc10

        \font\twelvecp=cmcsc10 scaled\magstep1
        
        \font\sixrm=cmr6
        \font\fiverm=cmr5

        \font\eightsy=cmsy8
        \font\sixsy=cmsy6
        \font\eightsl=cmsl8
        \font\sixbf=cmbx6

	 at10pt	
	\font\twelvehelvbold=phvb at12pt
	 at14pt
	\font\sixteenhelvbold=phvb at16pt

\def\noblackbox{\overfullrule=0pt}
\noblackbox

\def\eightpoint{
\def\rm{\fam0\eightrm}
\textfont0=\eightrm \scriptfont0=\sixrm \scriptscriptfont0=\fiverm
\textfont1=\eightmi  \scriptfont1=\sixmi  \scriptscriptfont1=\fivemi
\textfont2=\eightsy \scriptfont2=\sixsy \scriptscriptfont2=\fivesy
\textfont3=\tenex   \scriptfont3=\tenex \scriptscriptfont3=\tenex
\textfont\itfam=\eightit \def\it{\fam\itfam\eightit}
\textfont\slfam=\eightsl \def\sl{\fam\slfam\eightsl}
\textfont\ttfam=\eighttt \def\tt{\fam\ttfam\eighttt}
\textfont\bffam=\eightbf \scriptfont\bffam=\sixbf 
                         \scriptscriptfont\bffam=\fivebf
                         \def\bf{\fam\bffam\eightbf}
\normalbaselineskip=10pt}

\newtoks\headtext
\headline={\ifnum\pageno=1\hfill\else
	\ifodd\pageno{\eightcp\the\headtext}{ }\dotfill{ }{\old\folio}
	\else{\old\folio}{ }\dotfill{ }{\eightcp\the\headtext}\fi
	\fi}
\def\makeheadline{\vbox to 0pt{\vss\noindent\the\headline\break
\hbox to\hsize{\hfill}}
        \vskip2\baselineskip}
\newcount\infootnote
\infootnote=0
\newcount\footnotecount
\footnotecount=1
\def\foot#1{\infootnote=1
\footnote{${}^{\the\footnotecount}$}{\vtop{\baselineskip=.75\baselineskip
\advance\hsize by
-\parindent{\eightpoint\rm\hskip-\parindent
#1}\hfill\vskip\parskip}}\infootnote=0\global\advance\footnotecount by
1}
\newcount\refcount
\refcount=1
\newwrite\refwrite
\def\oldsize{\ifnum\infootnote=1\xold\else\old\fi}
\def\ref#1#2{
	\def#1{{{\oldsize\the\refcount}}\ifnum\the\refcount=1\immediate\openout\refwrite=\jobname.refs\fi\immediate\write\refwrite{\item{[{\xold\the\refcount}]} 
	#2\hfill\par\vskip-2pt}\xdef#1{{\noexpand\oldsize\the\refcount}}\global\advance\refcount by 1}
	}
\def\refout{\eightpoint\catcode`\@=11
        \xrm\immediate\closeout\refwrite
        \vskip2\baselineskip
        {\noindent\twelvecp References}\hfill\vskip\baselineskip
        \baselineskip=.75\baselineskip
        \input\jobname.refs
        \baselineskip=4\baselineskip \divide\baselineskip by 3
        \catcode`\@=\active\rm}

\def\skipref#1{\hbox to15pt{\phantom{#1}\hfill}\hskip-15pt}

\def\hepth#1{\href{http://xxx.lanl.gov/abs/hep-th/#1}{arXiv:\allowbreak
hep-th\slash{\xold#1}}}

\def\arxiv#1#2{\href{http://arxiv.org/abs/#1.#2}{arXiv:\allowbreak
{\xold#1}.{\xold#2}}} 
\def\jhep#1#2#3#4{\href{http://jhep.sissa.it/stdsearch?paper=#2\%28#3\%29#4}{J. High Energy Phys. {\xbold #1#2} ({\xold#3}) {\xold#4}}}

\def\CQG#1#2#3{Class. Quantum Grav. {\xbold#1} ({\xold#2}) {\xold#3}}

\def\JPA#1#2#3{J. Phys. {\xbf A}{\xbold#1} ({\xold#2}) {\xold#3}}
\def\LMP#1#2#3{Lett. Math. Phys. {\xbold#1} ({\xold#2}) {\xold#3}}
\def\MPLA#1#2#3{Mod. Phys. Lett. {\xbf A}{\xbold#1} ({\xold#2}) {\xold#3}}

\def\NPB#1#2#3{Nucl. Phys. {\xbf B}{\xbold#1} ({\xold#2}) {\xold#3}}

\def\PLB#1#2#3{Phys. Lett. {\xbf B}{\xbold#1} ({\xold#2}) {\xold#3}}
\def\PR#1#2#3{Phys. Rept. {\xbold#1} ({\xold#2}) {\xold#3}}
\def\PRD#1#2#3{Phys. Rev. {\xbf D}{\xbold#1} ({\xold#2}) {\xold#3}}
\def\PRL#1#2#3{Phys. Rev. Lett. {\xbold#1} ({\xold#2}) {\xold#3}}

\newcount\sectioncount
\sectioncount=0
\def\section#1#2{\global\eqcount=0
	\global\subsectioncount=0
        \global\advance\sectioncount by 1
	\ifnum\sectioncount>1
	        \vskip2\baselineskip
	\fi
\line{\twelvecp\the\sectioncount. #2\hfill}
       \vskip.5\baselineskip\noindent
        \xdef#1{{\old\the\sectioncount}}}
\newcount\subsectioncount
\def\subsection#1#2{\global\advance\subsectioncount by 1
\vskip.75\baselineskip\noindent\line{\tencp\the\sectioncount.\the\subsectioncount. #2\hfill}\nobreak\vskip.4\baselineskip\nobreak\noindent\xdef#1{{\old\the\sectioncount}.{\old\the\subsectioncount}}}
\def\immediatesubsection#1#2{\global\advance\subsectioncount by 1
\vskip-\baselineskip\noindent
\line{\tencp\the\sectioncount.\the\subsectioncount. #2\hfill}
	\vskip.5\baselineskip\noindent
	\xdef#1{{\old\the\sectioncount}.{\old\the\subsectioncount}}}
\newcount\subsubsectioncount
\def\subsubsection#1#2{\global\advance\subsubsectioncount by 1
\vskip.75\baselineskip\noindent\line{\tencp\the\sectioncount.\the\subsectioncount.\the\subsubsectioncount. #2\hfill}\nobreak\vskip.4\baselineskip\nobreak\noindent\xdef#1{{\old\the\sectioncount}.{\old\the\subsectioncount}.{\old\the\subsubsectioncount}}}
\newcount\appendixcount
\appendixcount=0
\def\appendix#1{\global\eqcount=0
        \global\advance\appendixcount by 1
        \vskip2\baselineskip\noindent
        \ifnum\the\appendixcount=1
        \hbox{\twelvecp Appendix A: #1\hfill}\vskip\baselineskip\noindent\fi
    \ifnum\the\appendixcount=2
        \hbox{\twelvecp Appendix B: #1\hfill}\vskip\baselineskip\noindent\fi
    \ifnum\the\appendixcount=3
        \hbox{\twelvecp Appendix C: #1\hfill}\vskip\baselineskip\noindent\fi}
\def\acknowledgements{\vskip2\baselineskip\noindent
        \underbar{\it Acknowledgements:}\ }
\newcount\eqcount
\eqcount=0
\def\Eqn#1{\global\advance\eqcount by 1
\ifnum\the\sectioncount=0
	\xdef#1{{\noexpand\oldsize\the\eqcount}}
	\eqno({\oldstyle\the\eqcount})
\else
        \ifnum\the\appendixcount=0
\xdef#1{{\noexpand\oldsize\the\sectioncount}.{\noexpand\oldsize\the\eqcount}}
                \eqno({\oldstyle\the\sectioncount}.{\oldstyle\the\eqcount})\fi
        \ifnum\the\appendixcount=1
	        \xdef#1{{\noexpand\oldstyle A}.{\noexpand\oldstyle\the\eqcount}}
                \eqno({\oldstyle A}.{\oldstyle\the\eqcount})\fi
        \ifnum\the\appendixcount=2
	        \xdef#1{{\noexpand\oldstyle B}.{\noexpand\oldstyle\the\eqcount}}
                \eqno({\oldstyle B}.{\oldstyle\the\eqcount})\fi
        \ifnum\the\appendixcount=3
	        \xdef#1{{\noexpand\oldstyle C}.{\noexpand\oldstyle\the\eqcount}}
                \eqno({\oldstyle C}.{\oldstyle\the\eqcount})\fi
\fi}
\def\eqn{\global\advance\eqcount by 1
\ifnum\the\sectioncount=0
	\eqno({\oldstyle\the\eqcount})
\else
        \ifnum\the\appendixcount=0
                \eqno({\oldstyle\the\sectioncount}.{\oldstyle\the\eqcount})\fi
        \ifnum\the\appendixcount=1
                \eqno({\oldstyle A}.{\oldstyle\the\eqcount})\fi
        \ifnum\the\appendixcount=2
                \eqno({\oldstyle B}.{\oldstyle\the\eqcount})\fi
        \ifnum\the\appendixcount=3
                \eqno({\oldstyle C}.{\oldstyle\the\eqcount})\fi
\fi}
\def\multi{\global\advance\eqcount by 1}
\def\multieqn#1{({\oldstyle\the\sectioncount}.{\oldstyle\the\eqcount}\hbox{#1})}
\def\multiEqn#1#2{\xdef#1{{\oldstyle\the\sectioncount}.{\old\the\eqcount}#2}
        ({\oldstyle\the\sectioncount}.{\oldstyle\the\eqcount}\hbox{#2})}
\def\multiEqnAll#1{\xdef#1{{\oldstyle\the\sectioncount}.{\old\the\eqcount}}}
\newcount\tablecount
\tablecount=0
\def\Table#1#2{\global\advance\tablecount by 1
       \xdef#1{\the\tablecount}
       \vskip2\parskip
       \centerline{\it Table \the\tablecount: #2}
       \vskip2\parskip}
\newtoks\url
\def\Href#1#2{\catcode`\#=12\url={#1}\catcode`\#=\active#2}
\def\href#1#2{{#2}}

\parskip=3.5pt plus .3pt minus .3pt
\baselineskip=14pt plus .1pt minus .05pt
\lineskip=.5pt plus .05pt minus .05pt
\lineskiplimit=.5pt
\abovedisplayskip=18pt plus 4pt minus 2pt
\belowdisplayskip=\abovedisplayskip
\hsize=14cm
\vsize=19cm
\hoffset=1.5cm
\voffset=1.8cm
\frenchspacing
\footline={}
\raggedbottom

\newskip\origparindent
\origparindent=\parindent

\def\*{\partial}
\def\punkt{\,\,.}
\def\komma{\,\,,}

\def\={\!=\!}
\def\small#1{{\hbox{$#1$}}}

\def\fraction#1{\small{1\over#1}}
\def\fr{\fraction}
\def\Fraction#1#2{\small{#1\over#2}}
\def\Fr{\Fraction}

\def\eg{{\it e.g.}}

\def\ie{{\it i.e.}}

\def\nlni{\hfill\break}

\def\a{\alpha}
\def\b{\beta}
\def\d{\delta}
\def\e{\varepsilon}

\def\G{\Gamma}
\def\L{\Lambda}
\def\O{\Omega}

\def\RR{{\Bbb R}}

\catcode`@=11                                   
\catcode`\|=12                                  
\catcode`\&=4                                   

\newcount\ncols         \ncols=\z@              
\newcount\nrows         \nrows=\z@              
\newcount\curcol        \curcol=\z@             
     
\newdimen\thinsize      \thinsize=0.6pt         
\newdimen\thicksize     \thicksize=1.5pt        

\newif\iftableinfo      \tableinfotrue          
\newif\ifcentertables   \centertablestrue       
%
%
     
\let\plaincr=\cr                        
\let\plainspan=\span                    
\let\plaintab=&                         
\let\lparen=(                           
\let\NX=\noexpand                       

     
\def\ruledtable{\relax                          
    \@BeginRuledTable                           
    \@RuledTable}


\def\@BeginRuledTable{
   \ncols=0\nrows=0                             
   \begingroup                                  
    \offinterlineskip                           
    \def~{\phantom{0}}
    \def\span{\plainspan\omit\relax\colcount\plainspan}
    \let\cr=\crrule                             
    \let\CR=\crthick                            
    \let\nr=\crnorule                           
    \let\|=\Vb                                  
%
%
    \ifx\tablestrut\undefined\relax             
    \else\let\tstrut=\tablestrut\fi             
    \catcode`\|=13 \catcode`\&=13\relax         
    \TableActive                                
    \curcol=1                                   
%
%
    \ifdim\tablewidth>-\maxdimen\relax          %
      \edef\@Halign{\NX\halign to \NX\tablewidth\NX\bgroup\TablePreamble}%
      \tabskip=0pt plus 1fil                    
    \else                                       %
      \edef\@Halign{\NX\halign\NX\bgroup\TablePreamble}%
      \tabskip=0pt                              
    \fi                                         %
%
%
    \ifcentertables                             
       \ifhmode\vskip 0pt\fi                    
       \line\bgroup\hss                         
    \else\hbox\bgroup                           
    \fi}


\long\def\@RuledTable#1\endruledtable{
   \vrule width\thicksize                       
     \vbox{\@Halign                             
       \thickrule                               
       #1\relax                                 
       \tstrut                                  
       \plaincr\thickrule                       
     \egroup}
   \vrule width\thicksize                       
   \ifcentertables\hss\fi\egroup                
  \endgroup                                     
  \global\tablewidth=-\maxdimen                 
  \iftableinfo                                  
      \immediate\write16{[Nrows=\the\nrows, Ncols=\the\ncols]}%
   \fi}
     

\def\TablePreamble{
   \linecount                           
   \TableItem{####}
   \plaintab\plaintab                   
   \TableItem{####}
   \plaincr}


\def\@TableItem#1{
   \hfil\tablespace                             
   #1\relax                                     
   \tablespace\hfil                             
    }%

\def\@tableright#1{
   \hfil\tablespace\relax               
   #1\relax                             
   \tablespace\relax}

\def\@tableleft#1{
   \tablespace\relax                    
   #1\relax                             
   \tablespace\hfil}

\let\TableItem=\@TableItem              
     
\def\RightJustifyTables{\let\TableItem=\@tableright}
\def\LeftJustifyTables{\let\TableItem=\@tableleft}
\def\NoJustifyTables{\let\TableItem=\@TableItem}

\def\LooseTables{\let\tablespace=\quad}
\def\TightTables{\let\tablespace=\space}
\LooseTables                                    

%

\newdimen\tablewidth    \tablewidth=-\maxdimen  


\def\setRuledStrut{
   \dimen@=\baselineskip                        
   \advance\dimen@ by-\normalbaselineskip       
   \ifdim\dimen@<.5ex \dimen@=.5ex\fi           
   \setbox0=\hbox{\lparen}
   \dimen1=\dimen@ \advance\dimen1 by \ht0      
   \dimen2=\dimen@ \advance\dimen2 by \dp0      
   \def\tstrut{\vrule height\dimen1 depth\dimen2 width\z@}%
   }%

\def\tstrut{\vrule height 3.1ex depth 1.2ex width 0pt}


\def\bigitem#1{
   \setbox0=\hbox{#1}
   \dimen1 =\ht0 \dimen2 =\dp0                  
   \dimen@ =\baselines@ve                       
   \advance\dimen@ by-\normalbaselineskip       
   \ifdim\dimen@<.25ex \dimen@=.25ex\fi         
   \advance\dimen1 by \dimen@                   
   \advance\dimen2 by \dimen@                   
   \vrule height\dimen1 depth\dimen2 width\z@   
   \copy0}

     
%

     
\def\nextcolumn#1{
   \plaintab\omit#1\relax\colcount              
   \plaintab}
     
\def\tab{
   \nextcolumn{\relax}}


\def\vb{
   \nextcolumn{\vrule width\thinsize}}

\def\Vb{
   \nextcolumn{\vrule width\thicksize}}


     
{\catcode`\|=13 \let|0
 \catcode`\&=13 \let&0
 \gdef\TableActive{\let|=\vb \let&=\tab}%
}


\def\crrule{\relax                      
   \tstrut                              
   \plaincr\tablerule                   
  }%

\def\crthick{\relax                     
   \tstrut                              
   \plaincr\thickrule                   
  }%
     
\def\crnorule{\relax                    
   \tstrut                              
   \plaincr                             
   }%
   

     
\def\tablerule{\noalign{\hrule height\thinsize depth 0pt}}%
\def\thickrule{\noalign{\hrule height\thicksize depth 0pt}}%


%
%
%
     

\def\linecount{\relax\global\ncols=\curcol      
   \global\curcol=1                             
   \global\advance\nrows by 1\relax}
     
\def\colcount{\relax                            %
   \global\advance\curcol by 1\relax}


\newdimen\parasize      \parasize=4in           

%

%

\def\begintable{\relax                          
    \@BeginRuledTable                           
    \@begintable}

\long\def\@begintable#1\endtable{
   \@RuledTable#1\endruledtable}


\catcode`@=12                                   


\catcode`\@11\relax
\newif\ify@autoscale \y@autoscaletrue \def\Yautoscale#1{\ifnum #1=0
  \y@autoscalefalse\else\y@autoscaletrue\fi}
\newdimen\y@b@xdim
\newdimen\y@boxdim \y@boxdim=13pt
\def\Yboxdim#1{\y@autoscalefalse\y@boxdim=#1}
\newdimen\y@linethick    \y@linethick=.3pt
\def\Ylinethick#1{\y@linethick=#1}
\newskip\y@interspace \y@interspace=0ex plus 0.3ex
\def\Yinterspace#1{\y@interspace=#1}
\newif\ify@vcenter   \y@vcenterfalse
\def\Yvcentermath#1{\ifnum #1=0 \y@vcenterfalse\else\y@vcentertrue\fi}
\newif\ify@stdtext   \y@stdtextfalse
\def\Ystdtext#1{\ifnum #1=0 \y@stdtextfalse\else\y@stdtexttrue\fi}
\newif\ify@enable@skew   \y@enable@skewfalse
\expandafter\ifx\csname enableskew\endcsname\relax
 \y@enable@skewfalse \else \y@enable@skewtrue\fi
\def\y@vr{\vrule height0.8\y@b@xdim width\y@linethick depth 0.2\y@b@xdim}
\def\y@emptybox{\y@vr\hbox to \y@b@xdim{\hfil}}
\ify@enable@skew
 \def\y@abcbox#1{\if :#1\else
   \y@vr\hbox to \y@b@xdim{\hfil#1\hfil}\fi}
 \def\y@mathabcbox#1{\if :#1\else
   \y@vr\hbox to \y@b@xdim{\hfil$#1$\hfil}\fi}
\else
 \def\y@abcbox#1{\y@vr\hbox to \y@b@xdim{\hfil#1\hfil}}
 \def\y@mathabcbox#1{\y@vr\hbox to \y@b@xdim{\hfil$#1$\hfil}}
\fi
\def\y@setdim{%
  \ify@autoscale%
   \ifvoid1\else\typeout{Package youngtab: box1 not free! Expect an
     error!}\fi%
   \setbox1=\hbox{A}\y@b@xdim=1.6\ht1 \setbox1=\hbox{}\box1%
  \else\y@b@xdim=\y@boxdim \advance\y@b@xdim by -2\y@linethick
  \fi}
\newcount\y@counter
\newif\ify@islastarg
\def\y@lastargtest#1,#2 {\if\space #2 \y@islastargtrue
  \else\y@islastargfalse\fi}
\def\y@emptyboxes#1{\y@counter=#1\loop\ifnum\y@counter>0
  \advance\y@counter by -1 \y@emptybox\repeat}
\def\y@nelineemptyboxes#1{%
  \vbox{%
    \hrule height\y@linethick%
    \hbox{\y@emptyboxes{#1}\y@vr}
    \hrule height\y@linethick}\vskip-\y@linethick}
\def\yng(#1){%
  \y@setdim%
  \hskip\y@interspace%
  \ifmmode\ify@vcenter\vcenter\fi\fi{%
  \y@lastargtest#1,
  \vbox{\offinterlineskip
    \ify@islastarg
     \y@nelineemptyboxes{#1}
    \else
     \y@ungempty(#1)
    \fi}}\hskip\y@interspace}
\def\y@ungempty(#1,#2){%
  \y@nelineemptyboxes{#1}
  \y@lastargtest#2,
  \ify@islastarg
   \y@nelineemptyboxes{#2}
  \else
   \y@ungempty(#2)
  \fi}
\def\y@nelettertest#1#2. {\if\space #2 \y@islastargtrue
  \else\y@islastargfalse\fi}
\def\y@abcboxes#1#2.{%
  \ify@stdtext\y@abcbox#1\else\y@mathabcbox#1\fi%
  \y@nelettertest #2.
  \ify@islastarg\unskip%
   \ify@stdtext\y@abcbox{#2}\else\y@mathabcbox{#2}\fi%
  \else\y@abcboxes#2.\fi}
 \newdimen\y@full@b@xdim
 \newcount\y@m@veright@cnt
\ify@enable@skew
 \def\y@get@m@veright@cnt#1#2.{%
   \if :#1 \advance\y@m@veright@cnt by 1\y@get@m@veright@cnt#2.\fi}
 \let\y@setdim@=\y@setdim
 \def\y@setdim{%
   \y@setdim@ \y@full@b@xdim=\y@b@xdim
   \advance\y@full@b@xdim by 1\y@linethick}
 \def\y@m@veright@ifskew#1{
   \y@m@veright@cnt=0 \y@get@m@veright@cnt#1.
   \moveright \y@m@veright@cnt\y@full@b@xdim}
\else
 \def\y@m@veright@ifskew#1{}
\fi
\def\y@nelineabcboxes#1{%
  \y@nelettertest #1.
  \ify@islastarg
   \y@m@veright@ifskew{#1}
    \vbox{
      \hrule height\y@linethick%
      \hbox{\ify@stdtext\y@abcbox#1\else\y@mathabcbox#1\fi\y@vr}
      \hrule height\y@linethick}\vskip-\y@linethick
  \else
   \y@m@veright@ifskew{#1}
    \vbox{
      \hrule height\y@linethick%
      \hbox{\y@abcboxes #1.\y@vr}%
      \hrule height\y@linethick}\vskip-\y@linethick
  \fi}
\def\young(#1){%
  \y@setdim%
  \hskip\y@interspace%
  \y@lastargtest#1,
  \ifmmode\ify@vcenter\vcenter\fi\fi{%
  \vbox{\offinterlineskip
    \ify@islastarg\y@nelineabcboxes{#1}%
    \else\y@ungabc(#1)%
    \fi}}\hskip\y@interspace}
\def\y@ungabc(#1,#2){%
  \y@nelineabcboxes{#1}%
  \y@lastargtest#2,
  \ify@islastarg\y@nelineabcboxes{#2}%
  \else\y@ungabc(#2)%
  \fi}
\catcode`\@12\relax
 


\input diagrams

\def\L{\Lambda}

\def\O{\Omega}

\def\ol{\overline}

\def\textfrac#1#2{\raise .45ex\hbox{\the\scriptfont0 #1}\nobreak\hskip-1pt/\hskip-1pt\hbox{\the\scriptfont0 #2}}

\def\LL{{\cal L}}


\def\frac{\Fr}

\def\mathbb{\Bbb}

\def\Enn{E_{n(n)}}

\def\ZZ{{\Bbb Z}}

\def\smalladj{\hbox{\sixrm adj}}



\ref\CederwallUfoldbranes{M. Cederwall, {\xit ``M-branes on U-folds''},
in proceedings of 7th International Workshop ``Supersymmetries and
Quantum Symmetries'' Dubna, 2007 [\arxiv{0712}{4287}].}

\ref\BermanPerryGen{D.S. Berman and M.J. Perry, {\xit ``Generalised
geometry and M-theory''}, \jhep{11}{06}{2011}{074} [\arxiv{1008}{1763}].}    

\ref\BermanMusaevThompson{D.S. Berman, E.T. Musaev and D.C. Thompson,
{\xit ``Duality invariant M-theory: gaugings via Scherk--Schwarz
reduction}, \jhep{12}{10}{2012}{174} [\arxiv{1208}{0020}].}

\ref\UdualityMembranes{V. Bengtsson, M. Cederwall, H. Larsson and
B.E.W. Nilsson, {\xit ``U-duality covariant
membranes''}, \jhep{05}{02}{2005}{020} [\hepth{0406223}].}

\ref\ObersPiolineU{N.A. Obers and B. Pioline, {\xit ``U-duality and M-theory''},
\PR{318}{1999}{113}, 
\nlni [\hepth{9809039}].}

\ref\BermanGodazgarPerry{D.S. Berman, H. Godazgar and M.J. Perry,
{\xit ``SO(5,5) duality in M-theory and generalized geometry''},
\PLB{700}{2011}{65} [\arxiv{1103}{5733}].} 

\ref\BermanMusaevPerry{D.S. Berman, E.T. Musaev and M.J. Perry,
{\xit ``Boundary terms in generalized geometry and doubled field theory''},
\PLB{706}{2011}{228} [\arxiv{1110}{3097}].} 

\ref\BermanGodazgarGodazgarPerry{D.S. Berman, H. Godazgar, M. Godazgar  
and M.J. Perry,
{\xit ``The local symmetries of M-theory and their formulation in
generalised geometry''}, \jhep{12}{01}{2012}{012}
[\arxiv{1110}{3930}].} 

\ref\BermanGodazgarPerryWest{D.S. Berman, H. Godazgar, M.J. Perry and
P. West,
{\xit ``Duality invariant actions and generalised geometry''}, 
\jhep{12}{02}{2012}{108} [\arxiv{1111}{0459}].} 

\ref\CoimbraStricklandWaldram{A. Coimbra, C. Strickland-Constable and
D. Waldram, {\xit ``$E_{d(d)}\times\hbox{\eightbbb R}^+$ generalised geometry,
connections and M theory'' }, \arxiv{1112}{3989}.} 

\ref\CremmerPopeI{E. Cremmer, B. Julia, H. L\"u and C.N. Pope,
{\xit ``Dualisation of dualities. I.''}, \NPB{523}{1998}{73} [\hepth{9710119}].}

\ref\HullT{C.M. Hull, {\xit ``A geometry for non-geometric string
backgrounds''}, \jhep{05}{10}{2005}{065} [\hepth{0406102}].}

\ref\HullM{C.M. Hull, {\xit ``Generalised geometry for M-theory''},
\jhep{07}{07}{2007}{079} [\hepth{0701203}].}

\ref\HullDoubled{C.M. Hull, {\xit ``Doubled geometry and
T-folds''}, \jhep{07}{07}{2007}{080}
[\hepth{0605149}].}

\ref\HullTownsend{C.M. Hull and P.K. Townsend, {\xit ``Unity of
superstring dualities''}, \NPB{438}{1995}{109} [\hepth{9410167}].}

\ref\PalmkvistHierarchy{J. Palmkvist, {\xit ``Tensor hierarchies,
Borcherds algebras and $E_{11}$''}, \jhep{12}{02}{2012}{066}
[\arxiv{1110}{4892}].} 

\ref\deWitNicolaiSamtleben{B. de Wit, H. Nicolai and H. Samtleben,
{\xit ``Gauged supergravities, tensor hierarchies, and M-theory''},
\jhep{02}{08}{2008}{044} [\arxiv{0801}{1294}].}

\ref\deWitSamtleben{B. de Wit and H. Samtleben,
{\xit ``The end of the $p$-form hierarchy''},
\jhep{08}{08}{2008}{015} [\arxiv{0805}{4767}].}

\ref\CederwallJordanMech{M.~Cederwall, {\xit ``Jordan algebra
dynamics''}, \PLB{210}{1988}{169}.} 

\ref\BerkovitsNekrasovCharacter{N. Berkovits and N. Nekrasov, {\xit
    ``The character of pure spinors''}, \LMP{74}{2005}{75}
  [\hepth{0503075}].}

\ref\HitchinLectures{N. Hitchin, {``\xit Lectures on generalized
geometry''}, \arxiv{1010}{2526}.}

\ref\KoepsellNicolaiSamtleben{K. Koepsell, H. Nicolai and
H. Samtleben, {\xit ``On the Yangian $[Y(e_8)]$ quantum symmetry of
maximal supergravity in two dimensions''}, \jhep{99}{04}{1999}{023}
[\hepth{9903111}].}

\ref\HohmHullZwiebachI{O. Hohm, C.M. Hull and B. Zwiebach, {\xit ``Background
independent action for double field
theory''}, \jhep{10}{07}{2010}{016} [\arxiv{1003}{5027}].}

\ref\HohmHullZwiebachII{O. Hohm, C.M. Hull and B. Zwiebach, {\xit
``Generalized metric formulation of double field theory''},
\jhep{10}{08}{2010}{008} [\arxiv{1006}{4823}].} 

\ref\HohmZwiebach{O. Hohm and B. Zwiebach, {\xit ``On the Riemann
tensor in double field theory''}, \jhep{12}{05}{2012}{126}
[\arxiv{1112}{5296}].} 

\ref\WestEEleven{P. West, {\xit ``$E_{11}$ and M theory''},
\CQG{18}{2001}{4443} [\hepth{0104081}].}

\ref\AndriotLarforsLustPatalong{D. Andriot, M. Larfors, D. L\"ust and
P. Patalong, {\xit ``A ten-dimensional action for non-geometric
fluxes''}, \jhep{11}{09}{2011}{134} [\arxiv{1106}{4015}].}

\ref\AndriotHohmLarforsLustPatalongI{D. Andriot, O. Hohm, M. Larfors,
D. L\"ust and 
P. Patalong, {\xit ``A geometric action for non-geometric
fluxes''}, \PRL{108}{2012}{261602} [\arxiv{1202}{3060}].}

\ref\AndriotHohmLarforsLustPatalongII{D. Andriot, O. Hohm, M. Larfors,
D. L\"ust and 
P. Patalong, {\xit ``Non-geometric fluxes in supergravity and double
field theory''}, Fortsch. Phys. {\xbold60} (2012) 1150 [\arxiv{1204}{1979}].}

\ref\DamourHenneauxNicolai{T. Damour, M. Henneaux and H. Nicolai,
{\xit ``Cosmological billiards''}, \CQG{20}{2003}{R145} [\hepth{0212256}].}

\ref\DamourNicolai{T. Damour and H. Nicolai, 
{\xit ``Symmetries, singularities and the de-emergence of space''},
\arxiv{0705}{2643}.}

\ref\EHTP{F. Englert, L. Houart, A. Taormina and P. West,
{\xit ``The symmetry of M theories''},
\jhep{03}{09}{2003}{020}2003 [\hepth{0304206}].}

\ref\PachecoWaldram{P.P. Pacheco and D. Waldram, {\xit ``M-theory,
exceptional generalised geometry and superpotentials''},
\jhep{08}{09}{2008}{123} [\arxiv{0804}{1362}].}

\ref\DamourHenneauxNicolaiII{T. Damour, M. Henneaux and H. Nicolai,
{\xit ``$E_{10}$ and a 'small tension expansion' of M theory''},
\PRL{89}{2002}{221601} [\hepth{0207267}].}

\ref\KleinschmidtNicolai{A. Kleinschmidt and H. Nicolai, {\xit
``$E_{10}$ and $SO(9,9)$ invariant supergravity''},
\jhep{04}{07}{2004}{041} [\hepth{0407101}].}

\ref\WestII{P.C. West, {\xit ``$E_{11}$, $SL(32)$ and central charges''},
\PLB{575}{2003}{333} [\hepth{0307098}].}

\ref\KleinschmidtWest{A. Kleinschmidt and P.C. West, {\xit
``Representations of $G^{+++}$ and the r\^ole of space-time''},
\jhep{04}{02}{2004}{033} [\hepth{0312247}].}

\ref\WestIII{P.C. West, {\xit ``$E_{11}$ origin of brane charges and
U-duality multiplets''}, \jhep{04}{08}{2004}{052} [\hepth{0406150}].}

\ref\PiolineWaldron{B. Pioline and A. Waldron, {\xit ``The automorphic
membrane''}, \jhep{04}{06}{2004}{009} [\hepth{0404018}].}

\ref\WestBPS{P.C. West, {\xit ``Generalised BPS conditions''},
\arxiv{1208}{3397}.}

\ref\BermanCederwallKleinschmidtThompson{D.S. Berman, M. Cederwall,
A. Kleinschmidt and D.C. Thompson, {\xit ``The gauge structure of
generalised diffeomorphisms''}, \jhep{13}{01}{2013}{64} [\arxiv{1208}{5884}].}

\ref\PalmkvistBorcherds{J. Palmkvist, {\xit ``Borcherds and Kac--Moody
  extensions of simple finite-dimensional Lie algebras''}, \arxiv{1203}{5107}.}

\ref\ParkSuh{J.-H. Park and Y. Suh, {\xit ``U-geometry: SL(5)''}, \arxiv{1302}{1652}.}

\ref\CederwallMinimalExcMult{M. Cederwall, {\xit ``Non-gravitational 
exceptional supermultiplets}, \arxiv{1302}{6737}.} 

\ref\PalmkvistDual{J. Palmkvist, work in progress.}

\ref\CederwallPalmkvistSerre{M. Cederwall and J. Palmkvist, {\xit
    ``Serre relations, constraints and partition functions''}, to appear.}

\ref\CoimbraStricklandWaldramII{A. Coimbra, C. Strickland-Constable and
D. Waldram, {\xit ``Supergravity as generalised geometry II:
$E_{d(d)}\times\hbox{\eightbbb R}^+$ and M theory'' }, \arxiv{1212}{1586}.}  

\ref\HohmZwiebachGeometry{O. Hohm and B. Zwiebach, {\xit ``Towards an
invariant geometry of double field theory''}, \arxiv{1212}{1736}.} 

\ref\JeonLeeParkRR{I. Jeon, K. Lee and J.-H. Park, {\xit
``Ramond--Ramond cohomology and O(D,D) T-duality''},
\jhep{12}{09}{2012}{079} [\arxiv{1206}{3478}].} 

\ref\PureSGI{M. Cederwall, {\xit ``Towards a manifestly supersymmetric
    action for D=11 supergravity''}, \jhep{10}{01}{2010}{117}
    [\arxiv{0912}{1814}].}  

\ref\PureSGII{M. Cederwall, 
{\xit ``D=11 supergravity with manifest supersymmetry''},
    \MPLA{25}{2010}{3201} [\arxiv{1001}{0112}].}

\ref\HohmZwiebachLarge{O. Hohm and B. Zwiebach, {\xit ``Large gauge
transformations in double field theory''}, \jhep{13}{02}{2013}{075}
[\arxiv{1207}{4198}].} 

\ref\CremmerLuPopeStelle{E. Cremmer, H. L\"u, C.N. Pope and
K.S. Stelle, {\xit ``Spectrum-generating symmetries for BPS solitons''},
\NPB{520}{1998}{132} [\hepth{9707207}].}

\ref\JeonLeeParkI{I. Jeon, K. Lee and J.-H. Park, {\xit ``Differential
geometry with a projection: Application to double field theory''},
\jhep{11}{04}{2011}{014} [\arxiv{1011}{1324}].}

\ref\JeonLeeParkII{I. Jeon, K. Lee and J.-H. Park, {\xit ``Stringy
differential geometry, beyond Riemann''}, 
\PRD{84}{2011}{044022} [\arxiv{1105}{6294}].}

\ref\JeonLeeParkIII{I. Jeon, K. Lee and J.-H. Park, {\xit
``Supersymmetric double field theory: stringy reformulation of supergravity''},
\PRD{85}{2012}{081501} [\arxiv{1112}{0069}].}

\ref\HohmKwak{O. Hohm and S.K. Kwak, {\xit ``$N=1$ supersymmetric
double field theory''}, \jhep{12}{03}{2012}{080} [\arxiv{1111}{7293}].}

\ref\HohmKwakFrame{O. Hohm and S.K. Kwak, {\xit ``Frame-like geometry
of double field theory''}, \JPA{44}{2011}{085404} [\arxiv{1011}{4101}].}

\ref\HohmKwakZwiebachI{O. Hohm, S.K. Kwak and B. Zwiebach, {\xit
``Unification of type II strings and T-duality''},
\PRL{107}{2011}{171603} [\arxiv{1106}{5452}].}

\ref\HohmKwakZwiebachII{O. Hohm, S.K. Kwak and B. Zwiebach, {\xit
``Double field theory of type II strings''}, \jhep{11}{09}{2011}{013}
[\arxiv{1107}{0008}].} 

\ref\Hillmann{C. Hillmann, {\xit ``Generalized $E_{7(7)}$ coset
dynamics and $D=11$ supergravity''}, \jhep{09}{03}{2009}{135}
[\arxiv{0901}{1581}].}

\ref\AldazabalGranaMarquesRosabal{G. Aldazabal, M. Gra\~na,
D. Marqu\'es and J.A. Rosabal, {\xit ``Extended geometry and gauged
maximal supergravity''}, \arxiv{1302}{5419}.}


\headtext={M. Cederwall, J. Edlund, A. Karlsson: 
``Exceptional geometry and tensor fields''}

\line{
\epsfxsize=18mm
\epsffile{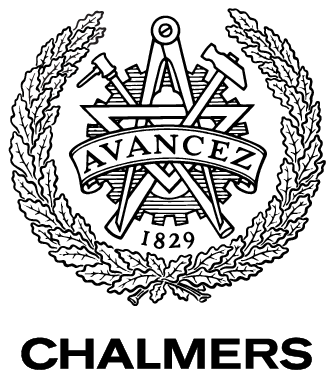}
\hfill}
\vskip-12mm
\line{\hfill Gothenburg preprint}
\line{\hfill February, {\old2013}}
\line{\hrulefill}

\vfill
\vskip.5cm

\centerline{\sixteenhelvbold
Exceptional geometry and tensor fields}

\vfill

\centerline{\twelvehelvbold{Martin Cederwall, Joakim Edlund and Anna Karlsson}}

\vfill

\centerline{\xrm Fundamental Physics}
\centerline{\xrm Chalmers University of Technology}
\centerline{\xrm SE 412 96 Gothenburg, Sweden}

\vfill

{\narrower\noindent \underbar{Abstract:} We present a tensor
  calculus for exceptional generalised geometry. Expressions for connections,
  torsion and curvature are given a unified formulation for different
  exceptional groups $E_{n(n)}$. We then consider ``tensor gauge
  fields'' coupled to the exceptional generalised gravity. 
Many of the properties of forms on
  manifolds are carried over to these fields.  
\smallskip}
\vfill

\font\xxtt=cmtt6

\vtop{\baselineskip=.6\baselineskip\xxtt
\line{\hrulefill}
\catcode`\@=11
\line{email: martin.cederwall@chalmers.se, joaedl@student.chalmers.se, karann@chalmers.se\hfill}
\catcode`\@=\active
}

\eject

\def\textfrac#1#2{\raise .45ex\hbox{\the\scriptfont0 #1}\nobreak\hskip-1pt/\hskip-1pt\hbox{\the\scriptfont0 #2}}

\def\fg{{\frak g}}
\def\fh{{\frak h}}
\def\fe{{\frak e}}
\def\fk{{\frak k}}

\section\Introduction{Introduction}The dualities of string theory or
M-theory treat momenta and brane charges on an equal footing. By
generalising space-time to include directions conjugate to brane
charges, such symmetries can be made manifest, but obviously the
concept of geometry has to be modified. There has been considerable
progress in the understanding of such models recently, both in the
context of U-duality 
[\HullTownsend\skipref\CremmerLuPopeStelle\skipref\CremmerPopeI-\ObersPiolineU],
which is the main focus of the present 
paper, and T-duality. We refer to both types of theories as
``generalised geometry''; doubled geometry [\HitchinLectures\skipref\HullT\skipref\HullDoubled\skipref\HohmHullZwiebachI\skipref\HohmHullZwiebachII\skipref\HohmZwiebach\skipref\HohmZwiebachGeometry\skipref\AndriotLarforsLustPatalong\skipref\AndriotHohmLarforsLustPatalongI\skipref\AndriotHohmLarforsLustPatalongII\skipref\JeonLeeParkI\skipref\JeonLeeParkII\skipref\JeonLeeParkIII\skipref\HohmKwak\skipref\HohmKwakFrame\skipref\HohmKwakZwiebachI\skipref\HohmKwakZwiebachII-\HohmZwiebachLarge] in the case of
T-duality, and exceptional geometry [\HullM\skipref\Hillmann\skipref\BermanPerryGen\skipref\BermanGodazgarPerry\skipref\BermanMusaevPerry\skipref\BermanGodazgarGodazgarPerry\skipref\BermanGodazgarPerryWest\skipref\BermanMusaevThompson\skipref\PachecoWaldram\skipref\CoimbraStricklandWaldram\skipref\CoimbraStricklandWaldramII\skipref\BermanCederwallKleinschmidtThompson-\ParkSuh] in the case of U-duality.

Turning to the state of the subject of exceptional geometry, it has
been shown that it is possible to formulate the dynamics of a generalised
metric, parametrising a coset $G/H$ with $G=E_{n(n)}\times\RR^+$ and
$H$ its maximal compact subgroup, in a manner which respects
local symmetries, generalising and including diffeomorphisms
[\BermanPerryGen\skipref\BermanGodazgarPerry\skipref\BermanMusaevPerry\skipref\BermanGodazgarGodazgarPerry-\BermanGodazgarPerryWest,\CoimbraStricklandWaldram,\BermanCederwallKleinschmidtThompson].
There are also results on an underlying geometry and tensor formalism
[\CoimbraStricklandWaldram,\ParkSuh],
but the covariant tensor calculus has so far been limited to $n=4$
[\ParkSuh].

The purpose of the present paper is twofold. We give a
universal (\ie, valid for all $n\leq7$) version of exceptional 
geometry, and a tensor
formalism that agrees with the one given for $n=4$ [\ParkSuh] and
makes manifest the
symmetry of ref. [\CoimbraStricklandWaldram]. 
We also initiate an investigation of what may be thought of as
differential geometry on a generalised manifold. A sequence of $G$
modules, in many respect analogous to forms on ordinary manifolds, are
given, and we describe how they may accommodate tensor
(non-gravitational) gauge fields.

The paper is organised as follows. After some background on
exceptional geometry in Section {\old2}, we turn to the covariant
construction of the generalised geometry in terms of vielbeins,
connections and curvature in Sections {\old3}-{\old5}. Section {\old6}
deals with the 
dynamics of tensor fields coupled to generalised geometry. We
summarise and point out some interesting questions in the concluding
Section.
Some conventions are given in an Appendix.

\section\Preliminaries{Preliminaries on exceptional
  geometry}As mentioned in the Introduction, we are concerned with a
generalisation of geometry, where the traditional r\^ole of $GL(n)$ in
ordinary geometry is subsumed by the group $G=E_{n(n)}\times\RR^+$,
and that of the locally realised rotation group by the maximal compact
subgroup $H\subset G$.  

Generalised momenta transform in a module $\ol R_1$ of $G$. 
A central identity in generalised geometry is the section
condition. 
It states that bilinears in momenta projected on a certain
module of $G$, $\ol R_2$, vanish. Although this condition is
$G$-covariant, its solutions effectively single out $n$ directions on
which fields may depend. 

It is well known how to form a generalised Lie derivative,
governing the generalised diffeomorphisms, which effectively include
tensor gauge transformation in addition to ordinary diffeomorphisms. 
The generalised diffeomorphisms, acting on a vector,
take the form
$$
\LL_UV^M=L_UV^M+Y^{MN}{}_{PQ}\*_NU^PV^Q\Eqn\LieDerAnsatz
$$
($L_U$ being the ordinary Lie derivative),
which can be rewritten as
$$
\eqalign{
\LL_UV^M&=U^N\*_NV^M
-\a P_{(\smalladj)}{}^M{}_{N,}{}^P{}_Q\*_PU^QV^N+\b\*_NU^NV^M\cr
&=U^N\*_NV^M+Z^{MN}{}_{PQ}\*_NU^PV^Q
\komma\cr}
\Eqn\AlgebraTransf
$$
where $P_{(\smalladj)}$ projects on the adjoint of
$\Enn$ (we constrain the analysis to $n\geq4$, where this group is
simple). 
For $n\leq6$, the tensor $Y$ is proportional to the projection on
$R_2$,
$$
Y^{MN}{}_{PQ}=2(n-1)P_{(R_2)}^{MN}{}_{PQ}\komma\eqn
$$
and for $n=7$ it contains an additional antisymmetric term 
$\fr2\e^{MN}\e_{PQ}$.
The constants $\a_n$ 
take the values $3,4,6,12$ for $n=4,5,6,7$, respectively, while
$\b_n={1\over9-n}$.

The closure of the algebra of generalised diffeomorphisms relies on
certain identities involving the invariant tensor $Y$. The
simplest of these is the section condition itself,
$$
Y^{MN}{}_{PQ}\*_M\otimes\*_N=0\komma\Eqn\SectionCondition
$$
where the $\otimes$ sign signifies that the two derivatives may act on
any pair of fields.
Another important identity is the nonlinear relation
$$
\left(Y^{MN}{}_{TQ}Y^{TP}{}_{RS}-Y^{MN}{}_{RS}\delta^P_Q\right)
\*_{(N}\otimes\*_{P)}=0\komma\Eqn\NonLinOne
$$
which can also be written
$$
\left(Z^{MN}{}_{TQ}Z^{TP}{}_{RS}+Z^{MP}{}_{RQ}\d^N_S\right)
\*_{(N}\otimes\*_{P)}=0\punkt\Eqn\NonLinTwo
$$
Notice, that while eq. (\NonLinOne) manifests the $R_2$ and $\ol R_2$
projections of the index pairs ${}^{MN}$ and ${}_{RS}$, the form (\NonLinTwo)
manifests the $\fg$ projections in the pairs ${}^M{}_Q$ and
${}^P{}_R$.

The parameters of generalised diffeomorphisms come in $R_1$, and it
was demonstrated in ref. [\BermanCederwallKleinschmidtThompson] that
the infinite sequences $\{R_k\}$ are responsible for the reducibility
of the transformations. As we will see in Section {\old6}, part of the
sequence has many properties in common with forms in ordinary
geometry, which is how we will be able to use them for constructing
tensor fields. Before that is possible, we need to develop a tensor
formalism.

\vskip4\parskip
\ruledtable
 $n$ \|  $R_1$ & $R_2$ & $R_3$ & $R_4$ & $R_5$ \crthick
3 \|    $({\bf3},{\bf2})$   &  $(\overline{\bf3},{\bf1})$  &
$({\bf1},{\bf2})$  &   $({\bf3},{\bf1})$ 
                & $(\overline{\bf3},\overline{\bf2})$ \cr 
4 \| ${\bf10}$& $\overline{\bf5}$ & ${\bf5}$&
 $\overline{\bf10}$ &${\bf24}$\cr   
5 \|  $\bf16$ & $\bf10$ & $\overline{\bf16}$ & $\bf45$ \cr 
6 \|  $\bf27$ & $\overline{\bf27}$ & $\bf78$ & \cr
7 \| $\bf{56}$ & $\bf{133}$  
 \endruledtable
\Table\ReducibilityTable{A partial list of modules $R^{(n)}_k$.}

\section\Tensors{Tensors and connections}The property (\AlgebraTransf)
of the generalised Lie derivative on vectors ensures that it can be
defined on a tensor carrying an arbitrary number of indices in $R_1$
and $\ol R_1$, with the transformation
$$
\eqalign{
\LL_U W^{M_1\ldots M_p}{}_{N_1\ldots N_q}
&=U^P\*_PW^{M_1\ldots M_p}{}_{N_1\ldots N_q}\cr
&+\sum\limits_{i=1}^p Z^{M_iQ}{}_{RP}\*_QU^R
     W^{M_1\ldots M_{i-1}PM_{i+1}\ldots M_p}{}_{N_1\ldots N_q}\cr
&-\sum\limits_{i=1}^q Z^{PQ}{}_{RN_i}\*_QU^R
     W^{M_1\ldots M_p}{}_{N_1\ldots N_{i-1}PN_{i+1}\ldots
       N_q}\komma\cr
}\Eqn\TensorTranformation
$$
so that tensor products and contractions respect the tensorial
property.
 
Note that composition of tensors implies that the $\RR$-weight is
not freely assigned. Not any invariant
$\Enn$ tensor is a tensor under generalised diffeomorphisms. For
example, $E_6$ has an invariant tensor $c^{MNP}$. In order to be a
tensor under generalised diffeomorphisms it would need to carry 
total $\RR$
weight 3, if the weight of a vector is normalised to one. Otherwise it
becomes a tensor density. On the other hand, $c^{MNP}c_{QRS}$ is a tensor.

We will introduce an affine connection, $\G_{MN}{}^P$. As matrices
$(\G_M){}_N{}^P$, 
$\G_M$ are valued in the Lie algebra $\fg={\frak e}_{n(n)}\oplus\RR$. 
Note that this excludes any specific symmetry properties for the lower
indices.
Defining a
covariant derivative $D=\*+\G$, the transformation rule of the
connection should ensure that $D_MW^{\{N\}}{}_{\{P\}}$ is a tensor if 
$W^{\{N\}}{}_{\{P\}}$ is a
tensor.
We use the convention 
$$
\eqalign{
D_MV_N&=\*_MV_N+\G_{MN}{}^PV_P\komma\cr
D_MV^N&=\*_MV^N-\G_{MP}{}^NV^P\komma\cr}
\Eqn\CovDevConv
$$
with the obvious generalisation to arbitrary number of indices.

The covariant derivatives of
  eq. (\CovDevConv) are valid for tensors, \ie, for objects where each
  $R_1$ index is accompanied with a certain $\RR$-weight $w$, which we may
  normalise to $1$, and accordingly $-1$ for each $\ol R_1$
  index. This is not always an ideal way of describing modules. 
One may for example want to use invariant tensors of $E_{n(n)}$ which do
  not have weight zero. One example is the duality $R_k\leftrightarrow
  \ol R_{9-n-k}$. It may sometimes be more convenient to represent,
  say, $R_{8-n}=\ol R_1$ with one lower index instead of $8-n$ upper
  ones. This amounts to considering ``tensor densities'', by
  specifying $E_{n(n)}$ module and $\RR$-weight $w$. There is no acute
  need of distinguishing tensors and ``tensor densities'', and we will
  use the term ``tensor'' for both. 
The covariant derivatives (taking a tensor of weight $w$ to one of
weight $w-1$) on vectors and covectors, with natural
generalisation to arbitrary index structures, are
$$
\eqalign{
D_MW_N&=\*_MW_N+\G_{MN}{}^PW_P-\Fr{w+1}{|R_1|}\G_{MP}{}^PW_N\komma\cr
D_MV^N&=\*_MV^N-\G_{MP}{}^NV^P-\Fr{w-1}{|R_1|}\G_{MP}{}^PV^N\punkt\cr
}\Eqn\CovDevDensity
$$

Demanding that the covariant derivative takes tensors to tensors 
immediately leads to the transformation rule for the connection,
$$
\eqalign{
\delta_\xi\G_{MN}{}^P&=\LL_\xi\G_{MN}{}^P+Z^{PQ}{}_{RN}\*_M\*_Q\xi^R\cr
&=\LL_\xi\G_{MN}{}^P-\*_M\*_N\xi^P+Y^{PQ}{}_{RN}\*_M\*_Q\xi^R\punkt\cr
}\Eqn\GammaTransformation
$$

As mentioned, the generic $\Enn$ module for the affine
connection is $\ol R_1\otimes\fg$. Not all of
the irreducible components of $\G$ can appear in the inhomogeneous
terms of eq. (\GammaTransformation). Only the part occurring in
$(\vee^2\ol R_1\ominus\ol R_2)\otimes R_1$ will pick up
inhomogeneous transformation terms. 
We define:
\item{\qquad}{\it Torsion is defined as the irreducible
modules in the affine connection transforming homogeneously,
\ie, with the generalised Lie derivative.}
 
\noindent Defined in this covariant way, torsion can consistently be
set to zero. 

It is quite straightforward to verify that the overlap
$[\ol R_1\otimes\fg]\cap[(\vee^2\ol
R_1\ominus\ol R_2)\otimes R_1]$ generically consists of a small
module, which is $\ol R_1$, and a big module, which
is the largest module in the product of $\ol R_1$ and the
adjoint. The torsion module, which is the rest of $\G$,
consists of a small module $\ol R_1$ and a bigger one (reducible for
low $n$), which
turns out to coincide with $\ol R_{10-n}$\foot{It has been
  observed in ref. [\CoimbraStricklandWaldram] that this 
torsion module can be identified with the embedding tensor of gauged
supergravity. 
Work by Palmkvist [\PalmkvistDual] identifies a new class of algebras,
symmetric under $R_p\rightarrow\ol R_{9-n-p}$ where torsion appears
as $R_{-1}$.}.

\vskip4\parskip
\vtop{
\ruledtable
$n$\|torsion|non-torsion\crthick
$3$\|$2(\overline{\bf3},{\bf2})\oplus({\bf6},{\bf2})$
             |$(\overline{\bf3},{\bf2})\oplus(\overline{\bf3},{\bf4})
                      \oplus({\bf15},{\bf2})$\cr
$4$\|$\overline{\bf10}\oplus\overline{\bf15}\oplus{\bf40}$
             |$\overline{\bf10}\oplus\overline{\bf175}$\cr
$5$\|$\overline{\bf16}\oplus{\bf144}$
           |$\overline{\bf16}\oplus\overline{\bf560}$\cr
$6$\|$\overline{\bf27}\oplus{\bf351'}$
            |$\overline{\bf27}\oplus\overline{\bf1728}$\cr
$7$\|${\bf56}\oplus{\bf912}$|${\bf56}\oplus{\bf6480}$
\endruledtable
\Table\TorsionTable{Torsion and non-torsion part of the affine
  connection.}
}

We need explicit expressions for the torsion, or expressions that a
torsion-free connection satisfies. It turns out that
$$
T_{MN}{}^P=\G_{MN}{}^P+Z^{PQ}{}_{RN}\G_{QM}{}^R\Eqn\CovTransfCond
$$
transforms as a tensor. This is verified by direct insertion into the
transformation rule (\GammaTransformation) and use of the identity
(\NonLinTwo). 
A torsion-free connection obeys
$$
\G_{MN}{}^P+Z^{PQ}{}_{RN}\G_{QM}{}^R=0\komma\Eqn\TorsionFree
$$
or, equivalently, $2\G_{[MN]}{}^P+Y^{PQ}{}_{RN}\G_{QM}{}^R=0$.
Note that the result from ordinary geometry is recovered for $Y=0$.

It is straightforward to take a trace to determine which
combination of the two $\ol R_1$'s is
torsion and which is torsion-free.
Contracting eq. (\TorsionFree) with $\d^N_P$ and using
$Z^{MP}{}_{PN}={|R_1|\over9-n}\d^M_N$ shows that a torsion-free
connection satisfies
$$
\G_{MN}{}^N+\Fr{|R_1|}{9-n}\G_{NM}{}^N=0\punkt\Eqn\ROneTorsion
$$
On the other hand, contracting eq. (\TorsionFree) with $\d_P^M$ gives
$$
Y_{MN}{}^{QR}\G_{QR}{}^N=-2\G_{[NM]}{}^N
=-\bigl(1+\Fr{|R_1|}{9-n}\bigr)\G_{NM}{}^N\punkt
\Eqn\YGamma
$$
For $n<7$ this identity may be used to derive a ``stronger''
constraint. Since $Y_{MN}{}^{QR}\G_{QR}{}^P$ can only contain the $\ol
R_1$ part of a torsion-free connection\foot{Because $\ol
  R_2\otimes R_1$ does not contain the big torsion-free connection 
module. This is not
true for $n=7$, where $\ol R_2$ is the adjoint.}, it must be
proportional to $Y_{MN}{}^{PQ}\G_{RQ}{}^R$, and the proportionality
constant is determined from eq. (\YGamma). The resulting relation is
$$
Y_{MN}{}^{QR}\G_{QR}{}^P+Y_{MN}{}^{PQ}\G_{RQ}{}^R=0\punkt\Eqn\YGammaTwo
$$
This relation is useful for determining when covariant
derivatives are connection-free; see below.

The generalised Lie
derivative on a vector does not contain any non-homogeneously transforming
connection, if one replaces the naked derivatives with covariant ones. 
This is verified by replacing the derivatives in $\LL_UV$ of
eq. (\AlgebraTransf) with covariant derivatives and checking that the
connections come in the torsion combination of
eq. (\TorsionFree). This property was used as a definition of torsion
(equivalent to ours) in ref. [\CoimbraStricklandWaldram].

Eq. (\TorsionFree) contains the torsion modules in the
  connection. The actual torsion-free connection cannot be obtained
  simply by adding a multiple of $T_{MN}{}^P$ to $\G_{MN}{}^P$,
  since the different torsion modules take different eigenvalues under
$\G\rightarrow T$.

\section\Compatibility{Vielbeins and compatible connections}The
structure group $G=E_{n(n)}\times\RR^+$ has a locally realised
subgroup $H$, which in the signature we are using is the maximal
compact subgroup $H=K(E_{n(n)})$. We denote $R_1$ indices under $H$ by
$A,B,\ldots$

Consider a vielbein (frame field) $E_M{}^A$, which is a group element of
$\Enn\times\RR^+$. Locally it represents an element of the coset $G/H$,
so it should be considered modulo local $H$-transformations from the
right. It can be used to form a metric $G_{MN}=E_M{}^AE_N{}^B\d_{AB}$,
where $\d_{AB}$ is an $H$-invariant constant metric.

We want to impose that the vielbein 
is covariantly constant, when transported by a
covariant derivative containing both affine and spin connections:
$$
D_ME_N{}^A=\*_ME_N{}^A+\G_{MN}{}^PE_P{}^A
          -E_N{}^B\O_{MB}{}^A=0\punkt
\Eqn\ConstantVielbein
$$

We now want to examine to what extent the connections are determined
from the vanishing of torsion together with the compatibility
equation (\ConstantVielbein). The affine connection can be eliminated
from the equation by the use of the vanishing torsion condition --- this
simply amounts to forming a combination of eq. (\ConstantVielbein) 
that contains $\G$ through $T$ of
eq. (\CovTransfCond). The result is
$$
(D^{(\O)}_MEE^{-1})_N{}^P+Z^{PQ}{}_{RN}(D^{(\O)}_QEE^{-1})_M{}^R=0\punkt
\Eqn\OmegaCompat
$$
On the other hand, the spin connection can be eliminated by projecting
the compatibility equation on its $\fg/\fh$ part. 
Note that when we talk about the local subgroup $H$ we always mean
the one defined by the vielbein. The projection is easy, since after
lowering one index, the symmetric part of $\fg$ is $\fg/\fh$ and the
antisymmetric part $\fh$.
This leads to
$$
(E^{-1}D^{(\G)}_ME)_{(AB)}=0\komma\Eqn\GammaCompat
$$
or, equivalently, 
$$
D_MG_{NP}=\*_MG_{NP}+2\G_{M(NP)}=0\punkt\Eqn\GammaCompatTwo
$$

To analyse the compatibility equations for the spin connection
(\OmegaCompat) and 
the affine connection (\GammaCompat), one must decompose into
$H$-modules. One then finds that the content of eq. (\OmegaCompat),
which is identical to the torsion modules of Table \TorsionTable, 
is smaller than
the content of $\O$, which is $R_1\otimes\fh$. The missing module $\Sigma$ is
the ``big'' irreducible module in $R_1\otimes\fh$, \ie,
the $H$-module whose highest weight is the sum of the highest weights
of $R_1$ and $\fh$.
Similarly, the same result is obtained from the compatibility for the
affine connection, so there is always an undefined part (in the same
module) of a torsion-free compatible affine connection. 
This is summarised in the table below, whose content agrees with
ref. [\CoimbraStricklandWaldram].

\vskip4\parskip
\ruledtable
$n$|$H$|undetermined connection $\Sigma$\crthick
$4$|$SO(5)$|${\bf35}=(04)$\cr
$5$|$(Spin(5)\times Spin(5))/\ZZ_2$|$({\bf4},{\bf20})\oplus({\bf20},{\bf4})
           =(01)(03)\oplus(03)(01)$\cr
$6$|$USp(8)/\ZZ_2$|${\bf594}=(2100)$\cr
$7$|$SU(8)/\ZZ_2$|${\bf1280}\oplus\overline{\bf1280}
       =(1100001)\oplus(1000011)$
\endruledtable
\Table\UndefinedConnTable{The undetermined part of a compatible
  torsion-free connection}

This means, that if connection is not to represent independent
degrees of freedom, one should only introduce covariant derivatives
mapping between certain special pairs of modules. Consider two
modules $U$ and $V$ under $H$ (or its double cover), 
and let a covariant derivative map from one
to the other. This means that $R_1\otimes U\supset V$. We are then
only allowed to do this for pairs where at the same time
$\Sigma\otimes U\not\supset V$. Some such pairs (``spinor'' and
``gravitino'' modules) were discussed in
refs. [\CoimbraStricklandWaldram,\CoimbraStricklandWaldramII], 
and we will encounter other ones later.

A special case of such well-defined covariant derivatives consists of
situations where not only the $\Sigma$ part of a connection is absent, but
where connection is altogether absent, and a covariant derivative
equals an ordinary derivative. Such connection-free actions of
derivatives will be important for our description of tensor gauge
fields in Section {\old6}, but we will already at this point check what
the weight of a vector $W^M$ must be in order for the divergence
$D_MW^M$ to be connection-free. From eq. (\CovDevDensity) it follows that
$$
D_MW^M=\*_MW^M-\G_{MN}{}^MW^N+\Fr{w-1}{9-n}\G_{NM}{}^NW^M\punkt
\Eqn\DivergenceDensity
$$
The connection terms cancel for $w=10-n$, which can be expressed as 
$$
|G|^{-{9-n\over2|R_1|}}D_MV^M=\*_M\bigl(|G|^{-{9-n\over2|R_1|}}V^M\bigr)
\Eqn\PowerOfG
$$
for a vector of weight 1.
This result will have
bearing on any discussion on measures and partial integration.

At this point, we would also like to comment on the relation between the
present approach and the one used in a recent paper by Park and Suh
[\ParkSuh]. There, the affine connection is subject to precisely the
right number of constraints to make it uniquely determined from
compatibility. In addition to the torsion condition, this procedure
amounts to setting, by hand, the $\Sigma$ module in $\G$ to zero. The
resulting derivative with connection is then not fully covariant, but
will behave as such acting between certain modules, the pairs
described in the previous paragraph. We tend to prefer the present,
geometric description, which allows for connections to transform as
such (both with respect to generalised diffeomorphisms and local $H$
transformations). 

\vfill\eject

\section\Curvature{Curvature}We will now examine how curvature can be defined. 
We write the transformation rule (\GammaTransformation) 
for the affine connection as
$$
\Delta_\xi\G_{MN}{}^P
\equiv(\d_\xi-\LL_\xi)\G_{MN}{}^P=Z^{PQ}{}_{RN}\*_M\*_Q\xi^R
\komma\eqn
$$
in order to manifest the inhomogeneous term. Tensors are characterised
by $\Delta_\xi=0$. This leads to the
corresponding transformation of its derivative:
$$
\eqalign{
\Delta_\xi\*_M\G_{NP}{}^Q&=Z^{QR}{}_{SP}\*_M\*_N\*_R\xi^S\cr
&+\Delta_\xi\G_{MR}{}^{Q}\G_{NP}{}^{R}
-\Delta_\xi\G_{MN}{}^{R}\G_{RP}{}^{Q}
-\Delta_\xi\G_{MP}{}^{R}\G_{NR}{}^{Q}\punkt\cr
}
\eqn
$$
There are two possibilities to make the $\*^3\xi$ terms vanish ---
antisymmetrisation $[MN]$ or symmetrisation and projection on $\ol
R_2$.
We have not found any way of directly using the $\ol
R_2$ (although it will become clear below that it really is a specific
combination of the two possibilities that leads to a tensor). 
Antisymmetrisation gives
$$
\Delta_\xi\left(\*_{[M}\G_{N]P}{}^Q+\G_{[M|P|}{}^R\G_{N]R}{}^Q\right)
=-\Delta_\xi\G_{[MN]}{}^R\G_{RP}{}^Q
=\fr2Y^{RS}{}_{TN}\Delta_\xi\G_{SM}{}^T\G_{RP}{}^Q\komma\Eqn\DeltaXiDGamma
$$
where we have used the tensor property of the torsion of 
eq. (\CovTransfCond) in the last step.
This is a nice form that reduces to the covariant transformation of
the Riemann tensor for ordinary geometry ($Y=0$).
The middle step clearly shows why an attempt to construct a ``Riemann
tensor'' fails, when the torsion-free condition does not suffice to
set $\G_{[MN]}{}^P$ to zero.
If however the expression on the right hand side of
eq. (\DeltaXiDGamma) is contracted with $\d^N_Q$ and symmetrised in
$(MP)$, 
it can be written as
$\Delta_\xi(\fr4Y^{RS}{}_{TQ}\G_{SM}{}^T\G_{RP}{}^Q)$. Therefore,
$$
\eqalign{
R_{MN}&=\*_{(M}\G_{|P|N)}{}^P-\*_P\G_{(MN)}{}^P\cr
&+\G_{(MN)}{}^Q\G_{PQ}{}^P-\G_{P(M}{}^Q\G_{N)Q}{}^P
-\fr2Y^{PQ}{}_{RS}\G_{PM}{}^S\G_{QN}{}^R\cr
}\Eqn\CurvatureTensor
$$
transforms as a tensor.
If we restrict to vanishing torsion, the last term may be rewritten
using eq. (\TorsionFree),
and the curvature takes the form
$$
\eqalign{
R_{MN}&=\*_{(M}\G_{|P|N)}{}^P-\*_P\G_{(MN)}{}^P\cr
&+\G_{(MN)}{}^Q\G_{PQ}{}^P-\fr2\G_{PM}{}^Q\G_{QN}{}^P
      -\fr2\G_{P(M}{}^Q\G_{N)Q}{}^P\punkt\cr
}\eqn
$$

An alternative way of deriving curvature is to start from
the covariant constancy of the generalised vielbein, 
eq. (\ConstantVielbein). The procedure is to
act with one more covariant derivative, and use only combinations
where second derivatives on the vielbein are absent, due to either
antisymmetry or the section condition. The result (which of course is
zero) should be
expressible as the difference of two tensors, of which the one
expressed in terms of $\O$ should be manifestly a tensor, and the one
expressed in $\G$ manifestly invariant under local transformations in
$H$. Then the equality of the two expressions implies that each of them
enjoys the property manifest in the other.

Acting with a second derivative on eq. (\ConstantVielbein) gives
$$
\eqalign{
0&=\*_M\*_NE_P{}^A+\*_M\G_{NP}{}^QE_Q{}^A-E_P{}^B\*_M\O_{NB}{}^A\cr
&-(\G_N\G_M)_P{}^QE_Q{}^A-E_P{}^B(\O_M\O_N)_B{}^A
+2(\G_{(M}E\O_{N)})_P{}^A\punkt\cr}
\Eqn\CurvPrel
$$ 
Antisymmetrising in $[MN]$ gives
$$
\eqalign{
0&=(\*_{[M}\G_{N]}+\G_{[M}\G_{N]})_P{}^QE_Q{}^A\cr
&-E_P{}^B(\*_{[M}\O_{N]}+\O_{[M}\O_{N]})_B{}^A\komma\cr
}\eqn
$$
exactly as in ordinary geometry.
The expression $\*_{[M}\O_{N]}$ on the second line is however not a
tensor, since
$\G_{[MN]}{}^P$ is not torsion. One has to form some combination of
terms so that the $\G\O$ terms in eq. (\CurvPrel) combine with the
$\*\O$ terms into covariant derivatives $D^{(\G)}$. They can then be
converted into $\O$ using $D_MA_N=E_N{}^AD_MA_A$.
This can be achieved with one contraction of indices and
symmetrisation in the remaining two (as in the construction of the
curvature above)\foot{Hohm and Zwiebach manage to form a
  4-index tensor in the $O(d,d)$ situation, where one has access to an
$H$-invariant metric [\HohmZwiebachGeometry]. 
We do not see how that construction generalises to
the exceptional cases.}. The resulting curvature is identical to
the one given in eq. (\CurvatureTensor), and its expression in terms
of $\O$ is
$$
\eqalign{
R_{MN}&=E_{(M}{}^A\*_{N)}\O_{BA}{}^B
   -E_{(M}{}^AE_{N)}{}^BE_C{}^P\*_P\O_{AB}{}^C
   -\fr2Y^{PA}{}_{B(M}E_{N)}{}^C\*_P\O_{AC}{}^B\cr
&\qquad+\O_{(MN)}{}^A\O_{BA}{}^B-\O_{AM}{}^B\O_{BN}{}^A\cr
&\qquad-\fr2Y^{AB}{}_{C(M}\left(\O_{|AB}{}^D\O_{D|N)}{}^C
       +\O_{|A|N)}{}^D\O_{BD}{}^C\right)\punkt\cr
}\eqn
$$
(Here, we have used vanishing torsion and restricted the calculation
to $n\leq6$. We have also converted indices with the vielbein.)

We do not have a direct proof that $R_{MN}$ exhausts the possible
curvature tensors, although we suspect that this is the case. 
It is however clear that it is large enough to contain anything we
need. For example, $R_1$ contains a 2-form in $n$ dimensions, 
so there is enough room in
$R_{MN}$ for the modules of an ordinary Riemann tensor.

An important question is to what extent this curvature is defined in
terms of a vielbein. This especially concerns its projection on
$\fg/\fh$, since that part is a candidate for a ``Ricci'' or
``Einstein'' tensor, providing equations of motion for the geometry.
A variation of the curvature gives at hand that
$$
\d R_{MN}=D_{(M}\d\G_{|P|N)}{}^P-D_P\d\G_{(MN)}{}^P\punkt\Eqn\DeltaCurv
$$
There is nothing here that prevents the undefined module $\Sigma$ from
appearing in the second term. 
But if we consider the projection on $\fg/\fh$, we observe
that $(\fg/\fh)\otimes R_1\not\supset\Sigma$, 
so the variation of $R_{MN}$ does not contain the $\Sigma$ part of $\d\G$.
Thus, $R_{\{MN\}}$, the projection of $R_{MN}$ on $\fg/\fh$, is
well-defined, and can serve as a Ricci tensor\foot{The independence of
the $\Sigma$ part of $\G$ cannot be observed by simply entering an
expression for $\G$ in terms of its decomposition in $H$-modules into
eq. (\CurvatureTensor). Then the $\G\G$ part of the second
term would seems to contain $\Sigma$. One has to realise that the $H$
subgroup defined by the vielbein/metric is special; only for this
subgroup the covariant derivatives respect the decomposition into
$H$-modules. We have checked in a couple of examples ($n=4,5$) that an
explicit decomposition in $H$ modules yields no $\Sigma^2$ in the
$\hbox{\eightfrak g}/\hbox{\eightfrak h}$ part, 
but indeed terms linear in $\Sigma$.}.  

From this it is also clear that the singlet, the curvature scalar
$R=G^{MN}R_{MN}$ (which is part of $R_{\{MN\}}$), is well-defined in
terms of the metric. 

It is tempting to think of the curvature scalar as a Lagrangian for
generalised gravity, whose variation should give an Einstein
tensor. This of course has to rely on partial integration, since
$$
\d R=\d(G^{MN}R_{MN})=\d G^{MN}R_{MN}
+D_M(\d\G_N{}^{MN}-\d\G_N{}^{NM})\punkt\eqn
$$
The $D\d\G$ terms cannot be discarded unless the expression is
multiplied by a scalar density from the measure, and it follows from
eq. (\PowerOfG) that this density must have weight $9-n$. So, if the
Lagrangian density is
$$
{\cal L}=|G|^{-{9-n\over2|R_1|}}R\komma\eqn
$$
the equations of motion for $G_{MN}$, the generalised Einstein's
equations, become
$$
R_{\{MN\}}+\Fr{9-n}{2|R_1|}G_{MN}R=0\punkt\Eqn\EinsteinEq
$$
For pure generalised gravity, this is of course equivalent to
$R_{\{MN\}}=0$, but in presence of matter fields, as in the following
section, eq. (\EinsteinEq) provides the left hand side of the 
generalised Einstein's
equations.

We note that our density $|G|^{-{9-n\over2|R_1|}}$ agrees with the one
given in ref. [\ParkSuh] for $n=4$. There, the density is written as
``$M^{-1}$'', where $M$ is the determinant of a metric on the
fundamental ${\bf5}$ of $SL(5)$. We have $-{9-n\over2|R_1|}=-\fr4$, but our
$G_{MN}$ is a metric on the module ${\bf10}$. The double weight of $G$
and the double size of the determinant together account for the factor
4 compared to ref. [\ParkSuh].

\Yboxdim4pt
\Yvcentermath1

\section\TensorFields{Tensor fields}It is well known that the $k$-form
gauge fields in dimensionally reduced theories come in the modules
$R_k$ under the U-duality group. 
Here, we instead ask for the dynamics
in the ``internal'' directions, \ie, for the descriptions of fields in $R_k$
on a generalised manifold (at least locally). We need to be able to
describe gauge symmetry and field equations, as well as some counting
of degrees of freedom.
The resulting description provides the U-duality version of the spinor
of Ramond--Ramond fields for T-duality and double field theory
[\JeonLeeParkRR]. 

The sequences $\{R_k\}$ are symmetric under $R_k\leftrightarrow\ol
R_{9-n-k}$ (and the proper reassignment of $\RR$ weight), in analogy
with forms. When we occasionally talk about modules $R_k$ outside the
window $1\leq k\leq8-n$, which \eg\ are needed for the complete
reducibility, we will take the ones for $k\geq9-n$ to agree with the
ones given in ref. [\BermanCederwallKleinschmidtThompson], which
agrees with the positive levels of a Borcherds algebra
[\PalmkvistBorcherds] (the precise reason for this will be the subject
of a future publication [\CederwallPalmkvistSerre]). For $k\leq0$, we
will assume that the symmetry around $k={9-n\over2}$ remains.
Seen as objects with $k$ upper indices, entities $F^{M_1\ldots M_k}$
in $R_k$ are in general neither totally antisymmetric nor symmetric,
but have mixed symmetry. $R_2$ is always symmetric, but already $R_3$
is a module of mixed symmetry \lower1.5pt\hbox{\yng(2,1)}.

In ref. [\BermanCederwallKleinschmidtThompson] it was shown how the
$R_k$'s arise as an infinite sequence of ghosts related to the
generalised diffeomorphisms and its reducibility. An essential
property is that a derivative, $\*:\,R_k\rightarrow R_{k-1}$, is
nilpotent, so the sequence forms a complex. 
With this knowledge, it seems natural that the same modules should be
responsible for gauge transformations of tensor fields (and their
reducibilities). 

We will now proceed to show that the sequence of modules 
$\{R_k\}_{k=1}^{8-n}$ in many respects plays
a r\^ole similar to that of forms on an ordinary manifold. 
An important piece of information is to what extent the affine
connection takes part in the covariantised operation 
$D:\,R_k\rightarrow R_{k-1}$.
Ideally, we would want connection to be absent, and ``$D=\*$'', 
in analogy with the
situation for the exterior derivative on forms.

It turns out that the derivative from $R_k$ to $R_{k-1}$ is
connection-free for $2\leq k\leq8-n$. For some simple cases, like
$R_2\rightarrow R_1$ ($n\leq6$), it is straightforward to show:
$$
\eqalign{
D_N W^{MN}&=\*_NW^{MN}-\G_{NP}{}^MW^{PN}-\G_{NP}{}^NW^{MP}\cr
&=\*_NW^{MN}-\fr{2(n-1)}\left(Y^{NP}{}_{RS}\G_{NP}{}^M
                           +Y^{MP}{}_{RS}\G_{NP}{}^N\right)W^{RS}=0\komma\cr
}\eqn
$$
with the use of eq. (\YGammaTwo). For $R_3\rightarrow R_2$, the proof
is more involved, and relies on the hook (\lower1.5pt\hbox{\yng(2,1)})
property of $R_3$. For higher $k$ it is more convenient to use
$R_{9-n-k}=\ol R_k$ and to treat them as tensor densities. For
example, the covariant derivative from $\ol R_1$ with weight $w$ 
to $\ol R_2$ is ($n\leq6$)
$$
Y_{MN}{}^{PQ}D_PW_Q=Y_{MN}{}^{PQ}
\bigl(\*_PW_Q-\Fr{8-n-w}{9-n}\G_{RP}{}^RW_Q\bigr)\komma\eqn
$$
where eq. (\YGammaTwo) has been used again, showing that the
derivative $R_{8-n}\rightarrow R_{7-n}$ is connection-free ($n\leq6$).

However, it is obvious from direct inspection that $R_1\rightarrow
R_0$ and $R_{9-n}\rightarrow R_{8-n}$ contain connection.
Neither is it possible to make the complex finite by using singlets at
$k=0$ and $k=9-n$; the corresponding derivatives also contain
connection. These singlets actually both take the r\^ole one would
have wanted from the other: the derivative ${\bf1}\rightarrow\ol R_1$
is connection-free for weight 0, and the divergence
$R_1\rightarrow{\bf1}$, as we have seen, is connection-free when the
singlet has weight $9-n$. In some sense, it looks as though we
had an $(9-n)$-dimensional manifold, but with an exterior derivative 
``acting the
wrong way''. To some extent, it becomes clearer from the diagrams in
Appendix B what happens. They depict the action of an ordinary
derivative on the modules $R_k$ decomposed into $GL(n)$ modules. There
are always two sequences containing forms. All sequences are finite,
but the ones starting at $R_1$ (or lower) or ending at $R_{8-n}$ (or
higher) consist of the tensor product of a complex of forms with some
non-trivial $GL(n)$ module.

The problematic situation at the limits of the connection-free window
does not prevent us from describing gauge connections and their field
strengths within the window. It makes it more complicated to describe
a gauge field in $R_1$ (more about this below), and it seems to
obstruct a complete covariant description of the full reducibility of
the gauge transformations at any $k$. 

Consider a gauge field $A$ in $R_{k+1}$, $1\leq k\leq7-n$. It will
have a field strength $F=\*A$ in $R_k$. There is a gauge symmetry
$\d_\L A=\*\L$ with parameter $\L$ in $R_{k+2}$ 
and a Bianchi identity $\*F=0$ in $R_{k-1}$.
(For $k=7-n$ the above
discussion shows a difficulty with the covariance of the gauge
transformation, and similarly with the Bianchi identity for $k=1$. We
will for the moment ignore this issue.)

Given a metric, there is a natural duality operation, taking $F$ in $R_k$ to
$*F$ in $R_{9-n-k}$. This can be written in two ways (analogous to
lower or upper indices for ordinary forms). One is obtained by simply
lowering the $k$ indices with the metric. This results in a tensor in
$\ol R_k$ with weight $-k$. A tensor in $R_{9-n-k}$ has weight
$9-n-k$, so the weight has to be adjusted by an appropriate power of
$|G|$. The correct dual field strength is
$$
*F_{M_1\ldots M_k}=|G|^{-{9-n\over2|R_1|}}
G_{M_1N_1}\ldots G_{M_kN_k}F^{N_1\ldots N_k}\punkt\eqn
$$ 
The other way is to use an invariant tensor $\Sigma^{A_1\ldots
  A_{9-n}}$, which after conversion of indices with inverse vielbeins
becomes a tensor $\Sigma^{M_1\ldots M_{9-n}}$ and write
$$
*F^{M_{k+1}\ldots M_{9-n}}=\Sigma^{M_1\ldots M_{9-n}}
        G_{M_1N_1}\ldots G_{M_kN_k}F^{N_1\ldots N_k}\punkt\eqn
$$ 
The equation of motion for $A$ can now be written 
$$
\*{*}F=0\punkt\eqn
$$
Since only connection-free derivatives have been used for forming the
field strengths and the equations of motion, it is clear that there are
no problems with undefined connection. The metric enters only through
the dualisation. There is a duality symmetry under $k\rightarrow9-n-k$
exchanging equations of motion and Bianchi identities.
Again, we find that a Lagrangian density $F*F$ with weight $9-n$ is
necessary in order to make partial integration possible.

It may seem that it is problematic to use a gauge potential in $R_1$,
since the field strength would belong to $R_0$, which is outside the
connection-free window. For a number of reasons (one is the field content of
maximally supersymmetric generalised supergravity, see below) 
one would still like
to have potentials in $R_1$. Although we will leave the detailed
formulation to future work, we would like to argue that it is
meaningful to have such a potential. The argument is based on
dimensional reduction of generalised gravity. We will consider
linearised fields. The linearised degrees of freedom of generalised
gravity lie in $\fg/\fh$. Consider the decomposition under
``dimensional reduction'', \ie, when $n$ is lowered by 1. We drop the
singlet part, which is irrelevant for the argument, and do not
consider the weights of resulting modules. Let us denote the module
$\fe_{n(n)}/\fk(\fe_{n(n)})$ by $\phi_n$. Under dimensional reduction,
$\phi_n\rightarrow\phi_{n-1}\oplus R^{(n-1)}_1\oplus{\bf1}$. The $R_1$
in the lower-dimensional theory is a ``generalised graviphoton'',
whose dynamics is dictated by generalised gravity in the higher
dimension. We have not examined the details of this, but it clearly
shows that one can have fields in $R_1$. 

The following is also worth noticing about derivatives on $R_1$.
Taking a derivative of a field $A$ in $R_1$ gives
$D_Q A^R=\*_Q A^R - \G_{QM}{}^R A^M$
We can use the $Z$-tensor to pick out the $\fg$ part:
$$
Z^{PQ}{}_{RN} D_Q A^R=Z^{PQ}{}_{RN}(\*_Q A^R - \G_{QM}{}^R A^M)    
     =Z^{PQ}{}_{RN}\*_Q A^R + \G_{MN}{}^P A^M      \komma\eqn      
$$
where the the torsion-free property was used for the second term.
If the free index pair ${}_N{}^P$ is projected on $\fg/\fh$, only
well-defined connection enters. In addition, the $\fg/\fh$ part of
the compatibility equation (\GammaCompatTwo) tells us that the
$\fg/\fh$-valued part of a compatible $\G_M$ contains a $\*_M$ and
obeys the section condition. Therefore, even if the derivative 
$R_1\rightarrow\fg/\fh$ contains connection, a field strength 
$F=(DA)|_{\fg/\fh}$
 allows for a gauge invariance with parameter in $R_2$.
Such an invariance is expected, since $R^{(n)}_1\rightarrow
R^{(n-1)}_1\oplus R^{(n-1)}_2\oplus{\bf1}$ under dimensional reduction.

We would like to say some words about the counting of degrees of
freedom, both off-shell and on-shell. The models we are dealing with
are effectively euclidean field theories, so in a strict sense it is
not meaningful to talk about local on-shell degrees of freedom. What
we mean is the number of physical polarisations the on-shell fields
would carry, had the model been formulated with another real form of $G$
corresponding to Minkowski signature after solution of the section
condition. This gives numbers that are of practical use, especially
when it comes to supersymmetric models
[\CoimbraStricklandWaldramII,\CederwallMinimalExcMult] and 
matching of bosonic and fermionic degrees of freedom.

The counting of off-shell degrees of freedom is straightforward. It is
simply given by the number of field components subtracted with the
number of gauge parameters. Here, the infinite reducibility has to be
taken into account, and we thus know that the number of off-shell
degrees of freedom of a gauge field in $R_k$ is
$$
N_k=\sum\limits_{\ell=0}^{\infty}(-1)^\ell|R_{k+\ell}|\punkt\eqn
$$
Such sums are na\"\i vely divergent (the terms are alternating but
growing) but have a meaningful regularisation
[\BerkovitsNekrasovCharacter,\BermanCederwallKleinschmidtThompson]. Of
course, it is enough to perform the regularisation for $N_1$ and
calculated the finite difference. The result for $1\leq k\leq8-n$ is
$$
N^{(n)}_k=\cases{|R^{(n-1)}_k|+1   &$k=1\komma$\cr
                 |R^{(n-1)}_k|     &$2\leq k\leq8-n\punkt$\cr}
\eqn
$$
The numbers have the property $N^{(n)}_k=N^{(n)}_{10-n-k}$.

The on-shell number of degrees of freedom can safely be deduced
from the observation that all the fields on the $n$-dimensional
solution of the section condition are forms (ordinary massless tensor
fields). Therefore, the number of on-shell degrees of freedom 
of a field in $R^{(n)}_k$ is obtained as $|R^{(n-2)}_k|$. The number
of physical polarisations of a field is obtained by regarding the
``same'' field in ``two dimensions less'', just as the
counting goes for massless fields in Minkowski space.
Since $R^{(n-2)}_{k+1}=\ol R^{(n-2)}_{10-n-k}$, this counting agrees with
the dualisation from a potential for $F$ in $R^{(n)}_{k+1}$ to a
potential for $*F$ in $R^{(n)}_{10-n-k}$.

The counting has been tested on a number of non-gravitational
supermultiplets [\CederwallMinimalExcMult]. Here we will illustrate it
by counting the bosonic degrees of freedom in the maximal
generalised supergravity.
Fields will transform under the $SL(11-n)$ or $SO(1,10-n)$ ``R-symmetry''
of the ``reduced'' directions, and behave as forms under these. 
If one associates $R_k$ with a $k$-form for
$k=1,\ldots[{11-n\over2}]$, and asks for a selfduality for
$R_{11-n\over2}$ when $n$ is odd, the resulting counting is as
follows:

\vskip4\parskip
\ruledtable
$n$\|gen. gravity|scalar coset|$R_k$|total\crthick
$4$\|2|28|${7\choose1}\times3+{7\choose2}\times2+{7\choose3}\times1=98$|128\cr
$5$\|6|21|${6\choose1}\times6+{6\choose2}\times3+\fr2\times{6\choose3}\times2=101$|128\cr
$6$\|13|15|${5\choose1}\times10+{5\choose2}\times5=100$|128\cr
$7$\|24|10|${4\choose1}\times16+\fr2\times{4\choose2}\times10=94$|$\,\,$128
\endruledtable
\Table\OnShellCounting{Counting of bosonic degrees of freedom for
  maximal supersymmetry.}

\noindent Note that for $n=7$ also $R_2=R_{9-n}={\bf133}$, which we have not
discussed above, is needed. Maybe the dual of the well-defined
derivative $R_1\rightarrow\fg/\fh$ can be of use. 
The appearance of fields as forms in $R_k$
is well known. In the present context it can also be obtained from
dimensional reduction. We have already seen that the generalised
gravity on dimensional reduction gives rise to a generalised
graviphoton in $R_1$. The generic rule for tensor fields is that
$R^{(n)}_k$ gives rise to $R^{(n-1)}_k$ and $R^{(n-1)}_{k+1}$ (with an
extra singlet for $k=1$ and $k=8-n$), in close analogy to form
fields. This is how the binomial coefficients are sequentially built.

\section\Conclusions{Conclusions}We have presented a tensor calculus for
exceptional generalised geometry. This includes universal and covariant
expressions for connections and curvatures. Our analysis agrees with
ref. [\CoimbraStricklandWaldram], but has manifest covariance, and
with ref. [\ParkSuh] for $n=4$.
We have also given details on tensor gauge fields and their coupling
to exceptional geometry.
Some technical issues remain concerning the ``generalised
graviphoton'' field. 
Even if the local description in terms of a tensor calculus respecting
infinitesimal transformation now seems complete, important questions
concerning the concept of generalised manifolds remain open. Hohm and
Zwiebach have discussed the issue of exponentiating the Lie derivative
in double field theory to a large diffeomorphism, but there are many
remaining questions. An important one is to find an integration measure.

In ref. [\CoimbraStricklandWaldramII], minimal exceptional
supergravity was formulated. In an accompanying paper
[\CederwallMinimalExcMult] non-gravitational 
supermultiplets based on the tensor fields we present here were
constructed. 
Extended supergravity will demand
inclusion of such multiplets. 
It would be very interesting to investigate the possibility of
formulating such models as some generalised supergeometry. It is not
clear which set of modules will accompany the $R_k$'s in order to
build the correspondence to ``forms on superspace''.
Such a formulation, preferably in an off-shell version using pure
spinor techniques generalising refs. [\PureSGI,\PureSGII], 
could perhaps provide a
simultaneous manifestation of supersymmetry and U-duality. 

\vskip4\parskip
\noindent\underbar{Note added:} The paper [\AldazabalGranaMarquesRosabal],
which appeared on the completion of our work, specialises on $n=7$ and
has a substantial overlap with the present paper concerning the
geometric analysis.

\acknowledgements MC would like to thank Axel Kleinschmidt, Jakob
Palmkvist and David Berman for discussions.

\appendix{Notation}$G$ and $H$ denote throughout the paper the groups
$G=E_{n(n)}\times\RR^+$ and its compact subgroup $H=K(E_{n(n)})$, and
their Lie algebras (and adjoint modules) 
are written $\fg$ and $\fh$. For the complement to
$\fh$ in $\fg$ we use ``$\fg/\fh$'' (even if ``$\fg\ominus\fh$'' might
have been more correct). A projection of a 2-index object on $\fg/\fh$
is denoted by curly brackets: $\{MN\}$.

We use the notation $\vee$ for symmetrised tensor
product. The dimension of a module $R$ is denoted $|R|$. When a module in the
sequence $\{R_k\}$ carries an upper index, $R^{(n)}_k$, it refers to
$n$, the rank of the exceptional group.

\vfill\eject

\appendix{The action of a derivative among the $R_k$}Below 
are diagrams showing the action of a derivative fulfilling the
section condition on elements in $R_k$, $0\leq k\leq9-n$. The modules
are split into 
modules of $SL(n)\times\RR$. 
For $n=6,7$, there is an $SL(2)$
which is broken to $\RR$ by the solution of the section condition.

Note that there are always two lines containing the ordinary
$n$-dimensional forms. Other lines consist of the tensor product of
the forms by some non-trivial module. Such lines begin at $R_1$ and
end at $R_{8-n}$, and may be seen as responsible for the appearance of
connection.

\vskip2\parskip

\underbar{$n=4$:}
\diagram[height=2.4mm,width=5mm,tight]
{\bf\overline4}_1&&&&&&&&&&&&&&&{\bf\overline4}_1\\
&&&&&&&&&{\bf1}_{4/5}&&&&&\ldTo(3,3)\\
&&&{\bf4}_{3/5}&&&&&\ldTo(3,3)&&&&&\\
&&\ldTo(3,3)&&&&&&&&&&{\bf6}_{2/5}\\
&&&&&&{\bf\overline4}_{1/5}&&&&&\ldTo(3,3)\\
({\bf4}\otimes{\bf\overline4})_0&&&&&\ldTo(3,3)&&&&&
                      &&&&&({\bf4}\otimes{\bf\overline4})_0\\
&&&&&&&&&{\bf4}_{-1/5}&&&&&\ldTo(3,3)\\
&&&{\bf6}_{-2/5}&&&&&\ldTo(3,3)&&&&&\\
&&\ldTo(3,3)&&&&&&&&&&{\bf4}_{-3/5}\\
&&&&&&{\bf1}_{-4/5}\\
{\bf4}_{-1}&&&&&&&&&&&&&&&{\bf4}_{-1}
\enddiagram

\underbar{$n=5$:}
\diagram[height=3mm,width=5mm,tight]
&&&&&&&&&{\bf1}_{5/4}\\
{\bf\overline{10}}_1&&&&&&&&\ldTo(3,3)&&&&{\bf\overline{10}}_1\\
&&&{\bf5}_{3/4}&&&&&&&&\ldTo(3,3)\\
&&\ldTo(3,3)&&&&{\bf\overline5}_{1/2}&&&&\\
&&&&&\ldTo(3,3)&&&&{\bf10}_{1/4}\\
({\bf5}\otimes{\bf\overline5})_0&&&&&&&&\ldTo(3,3)
                          &&&&({\bf5}\otimes{\bf\overline5})_0\\
&&&{\bf\overline{10}}_{-1/4}&&&&&&&&\ldTo(3,3)\\
&&\ldTo(3,3)&&&&{\bf5}_{-1/2}&&&&\\
&&&&&\ldTo(3,3)&&&&{\bf\overline5}_{-3/4}\\
{\bf10}_{-1}&&&&&&&&&&&&{\bf10}_{-1}\\
&&&{\bf1}_{-5/4}
\enddiagram

\underbar{$n=6$:}
\diagram[height=6mm,width=7.5mm,tight]
{\bf1}_2&&&&&&{\bf1}_2\\
&&&&&\ldTo\\
{\bf20}_1&&{\bf6}_1&&{\bf\overline6}_1&&{\bf20}_1\\
&\ldTo&&\ldTo&&\ldTo\\
({\bf6}\otimes{\bf\overline6})_0&&{\bf\overline{15}}_0&&{\bf15}_0
     &&({\bf6}\otimes{\bf\overline6})_0\\
&\ldTo&&\ldTo&&\ldTo\\
{\bf20}_{-1}&&{\bf6}_{-1}&&{\bf\overline6}_{-1}&&{\bf20}_{-1}\\
&\ldTo&&&&\\
{\bf1}_{-2}&&&&&&{\bf1}_{-2}
\enddiagram

\underbar{$n=7$:}
\diagram[height=6mm,width=5mm,tight]
{\bf\overline7}_2&&&&&&{\bf\overline7}_2\\
&&&{\bf7}_{3/2}&&\ldTo(3,3)\\
{\bf35}_1&&\ldTo(3,3)&&&&{\bf35}_1\\
&&&{\bf\overline{21}}_{1/2}&&\ldTo(3,3)\\
({\bf7}\otimes{\bf\overline7})_0&&\ldTo(3,3)&&
             &&({\bf7}\otimes{\bf\overline7})_0\\
&&&{\bf21}_{-1/2}&&\ldTo(3,3)\\
{\bf\overline{35}}_{-1}&&\ldTo(3,3)&&&&{\bf\overline{35}}_{-1}\\
&&&{\bf\overline7}_{-3/2}&&\\
{\bf7}_{-2}&&&&&&{\bf7}_{-2}
\enddiagram

\refout
\end